\documentclass[12pt,twoside]{article}
\usepackage{amsmath,mathrsfs,bm,amssymb,color}

\usepackage[dvips]{graphicx}
\usepackage{latexsym}
\catcode`\@=11
\def\newsymbol#1#2#3#4#5{\let\next@\relax%
 \ifnum#2=\@ne\else%
 \ifnum#2=\tw@\let\next@\msyfam@\fi\fi%
 \mathchardef#1="#3\next@#4#5}
\def\mathhexbox@#1#2#3{\relax%
 \ifmmode\mathpalette{} {\m@th\mnnathchar"#1#2#3}
 \else\leavevmode\hbox{$\m@th\mathchar"#1#2#3$}\fi}
\font\tenmsy=msbm10
\font\sevenmsy=msbm7
\font\fivemsy=msbm5
\newfam\msyfam
\textfont\msyfam=\tenmsy
\scriptfont\msyfam=\sevenmsy
\scriptscriptfont\msyfam=\fivemsy
\edef\msyfam@{\hexnumber@\msyfam}
\def\mathbb #1{\fam\msyfam\relax#1}
\catcode`\@=\active

\newtheorem{theorem}{Theorem}[section]
\newtheorem{proposition}[theorem]{Proposition}
\newtheorem{lemma}[theorem]{Lemma}
\newtheorem{corollary}[theorem]{Corollary}
\newtheorem{definition}[theorem]{Definition}

\newtheorem{remark}[theorem]{Remark}

\newcommand{\Ebb}{{\mathbb  E}}

\newcommand{\bd}[1]{\begin{definition}\label{#1}}
\newcommand{\ed}{\end{definition}}

\renewcommand{\d}{\displaystyle}

\newcommand{\bl}[1]{\begin{lemma}\label{#1}}
\newcommand{\el}{\end{lemma}}
\newcommand{\bc}[1]{\begin{corollary}\label{#1}}
\newcommand{\ec}{\end{corollary}}
\newcommand{\bt}[1]{\begin{theorem}\label{#1}}
\newcommand{\et}{\end{theorem}}
\newcommand{\bp}[1]{\begin{proposition}\label{#1}}
\newcommand{\ep}{\end{proposition}}
\newcommand{\br}[1]{\begin{remark}\label{#1}}
\newcommand{\er}{\end{remark}}
\newcommand{\zz}{{\mathbb  Z}_2}

\newcommand{\eq}[1]{\begin{equation}\label{#1}}
\newcommand{\en}{\end{equation}}
\newcommand{\eqn}{\begin{eqnarray*}}
\newcommand{\enn}{\end{eqnarray*}}
\newcommand{\eqnn}{\begin{eqnarray}}
\newcommand{\ennn}{\end{eqnarray}}
\newcommand{\proof}{{\noindent \it Proof:\ }}
\newcommand{\qed}{\hfill {\bf qed}\par\medskip}

\newcommand{\bi}{\begin{description}}
\newcommand{\ei}{\end{description} }
\newcommand{\CC}{{{\mathbb  C}}}
\newcommand{\RR}{{\mathbb  R}}

\newcommand{\pro}[1]{(#1_t)_{t\geq0}}

\newcommand{\kak}[1]{(\ref{#1})}

\newcommand{\LR}{{L^2(\RR)}}
\newcommand{\LRM}{{L^2(\RR,d\mu)}}

\newcommand{\ms}{\mathscr }

\newcommand{\is}{\inf\sigma}

\newcommand{\ov}[1]{\overline{#1}}

\newcommand{\lk}{\left(}
\newcommand{\rk}{\right)}
\newcommand{\lkk}{\left\{}
\newcommand{\rkk}{\right\}}
\newcommand{\lkkk}{\left[}
\newcommand{\rkkk}{\right]}

\newcommand{\add}{a^{\dagger}}

\newcommand{\ww}{\frac{1}{\sqrt \omega}}
\newcommand{\w}{\sqrt\omega}

\newcommand{\PF}{{H_{\rm Rabi}}}

\newcommand{\gr}{\varphi_{\rm g}}

\def\bbbone{{\mathchoice {\rm 1\mskip-4mu l} {\rm 1\mskip-4mu l}
{\rm 1\mskip-4.5mu l} {\rm 1\mskip-5mu l}}}
\def\one{\bbbone}

\newcommand{\half}{\frac{1}{2}}
\newcommand{\han}{{1/2}}

\newcommand{\vvv}[1]{\left[
\!\!\!\begin{array}{c}#1\end{array}\!\!\!\right]}

\newcommand{\s}{\sigma}

\makeatletter
\@addtoreset{equation}{section}
\makeatother

\title
{\sc {Absence of Energy Level Crossing} \\ 
{for the Ground State Energy} \\ 
{of the Rabi Model}}

\author{
\small Masao Hirokawa \\
{\it \small Department of Mathematics, Okayama University} \\[-0.7ex]
{\it \small Okayama, 700-8530, Japan} \\[-0.7ex]
{\small {\tt hirokawa@math.okayama-u.ac.jp }} \\[0.3cm]
\small Fumio Hiroshima\\
{\small\it Faculty of Mathematics, Kyushu University}    \\[-0.7ex]
{\small\it 744 Motooka, Fukuoka, 819-0395, Japan}      \\[-0.7ex]
{\small  {\tt hiroshima@math.kyushu-u.ac.jp}}
\\[-0.7ex]}

\bigskip\bigskip\bigskip

\date{}

\begin{document}
\maketitle
\setlength{\baselineskip}{17pt}

\bigskip\bigskip\bigskip

\begin{abstract}
{The Hamiltonian of the Rabi model 
is considered. 
It is shown that the {ground state energy} of 
the Rabi Hamiltonian  is simple for all values of the coupling strength, 
which implies the ground state energy does not cross 
other energy}. 
\end{abstract}

\section{Introduction}

Cavity quantum electrodynamics has supplied us with 
stronger interaction than the standard quantum electrodynamics (QED) 
does \cite{HR-RBH,rbh01}.  
Experimental physicists usually demonstrate the interaction 
by coupling a two-level atom with a one-mode light (i.e., single-mode laser) 
in a mirror cavity (i.e., a mirror resonator). 
The region that the strong interaction in cavity QED 
belongs to is called the strong coupling regime. 
At the dawn of the $21$st century, 
the solid-state analogue of the strong interaction 
in a superconducting system was theoretically proposed 
in \cite{MSS,MB}, and it has been experimentally demonstrated 
in \cite{Chiorescu, Wallraff08,Wallraff04}. 
That is, the atom, the light, and the mirror resonator in cavity QED 
are respectively replaced by an artificial atom, 
a microwave, and a microwave resonator 
on a superconducting circuit. 
Here, the artificial atom is made by using 
a superconducting circuit based on some Josephson junctions. 
This replaced cavity QED is the so-called circuit QED. 
The circuit QED has been intensifying the coupling strength 
so that its region is beyond the strong coupling regime.  
This amazing region of the coupling 
strength between the artificial atom and the light 
is called the ultra-strong coupling regime 
in circuit QED \cite{DGS,ciuti,Mooij,Gross}. 
Then, experimental physicists have found some differences 
in physical phenomena between the two coupling regimes 
\cite{Mooij,Gross}. 
As one of the striking differences, there is the following. 
In the strong coupling regime as well as 
in the weak coupling regime, 
the Jaynes-Cummings (JC) model is useful to 
explain the experimental results \cite{HR-RBH,Wallraff08}. 
The Hamiltonian of the JC model is obtained 
by applying the so-called 
rotating wave approximation (RWA) to the Rabi Hamiltonian. 
On the other hand, in the ultra-strong coupling regime, 
the JC model does not work, and thus, 
we need a help of the Rabi model \cite{Mooij,Gross}. 
The current cutting-edge technology of circuit QED 
is begining to show us the division between 
the two {coupling regimes concretely. 
We are interested in how physics determines this division.}

To see a difference between the Rabi model and 
the JC model, in this letter 
we pay particular attention to 
the energy level crossing.  
Recently, Braak \cite{bra11} had given a mathematically intriguing 
expressions of the eigenenergies of the Rabi model. 
Then, 
the following questions arise 
and are problems of interest to us: 
(i) are there any energy level crossings among them? 
If so, (ii) how do they take place?
As is shown in \cite{hir09a,hir09b}, the ground state energy 
of the JC model consists of many energy level crossings 
as the coupling strength grows larger and larger. 
Namely, for the JC model the quantum phase transition 
in Rey's sense \cite{Rey09} takes place (see Fig\ref{fig:JC} below). 
 For the details on energy level crossing and quantum phase transition, 
see \cite{sach}.
On the other hand, as in our numerical computation in 
Fig.\ref{fig:Rabi}, 
we can conjecture that 
the ground state energy of the Rabi model has no 
energy level crossing. 
In this letter 
we prove this fact with the functional-integral method 
\cite{hl07,hil12,hhl12}
as a corollary of the fact stating that 
the ground state energy of the Rabi model is simple 
(i.e., the ground state is unique). 
It reveals us that it is in the ultra-strong coupling regime 
of circuit QED that there is a big qualitative difference 
as well as quantitative one between the Rabi model 
and the JC model.

\section{Rabi model}
\subsection{Definition}
Let $\sigma_x,\sigma_y,\sigma_z$ be the $2\times 2$ Pauli matrices:
\begin{align}
 \sigma_x= \begin{pmatrix} 0 & 1 \\ 1 & 0 \end{pmatrix},\quad
 \sigma_y = \begin{pmatrix} 0 & -i \\ i & 0 \end{pmatrix},\quad
 \sigma_z = \begin{pmatrix} 1 & 0 \\ 0 & -1 \end{pmatrix}.
\end{align}
In this letter we adopt the natural unit: 
$\hbar=1$. 
{The renormarized Hamiltonian 
of the Rabi model is defined 
as a self-adjoint operator by} 
\eq{rabi}
\PF={\s_z\Delta+\omega \add a+ g\s_x(a+\add)}
\en
on $\CC^2\otimes\LR$. 
Here {$\Delta>0$ and $\omega>0$ are respectively 
the atom transition frequency and the cavity resonance frequency, 
$g\in\RR$ stands for a coupling constant, 
and $a$ and $\add$ denote} the single mode bose annihilation and  creation operators satisfying $[a,\add]=1$. 
It is given by $a=(\ww \frac{d}{dx}+\w x)/\sqrt 2$ and $\add=(-\ww \frac{d}{dx}+\w x)/\sqrt 2$.
We are interested in studying {spectral properties of 
eigenenergies} of $\PF$, in particular 
crossing of the ground state energy. 

The {absence of} crossing 
can be derived from the simplicity of the ground state energy of $\PF$. 
We will construct a path integral representation of $e^{-t\PF}$ 
to show that the ground state energy is simple.  
This is a minor modification of recent papers \cite{hl07,hil12,hhl12}, where 
the Feynman-Kac type formula with spin is established.  
In particular the spin-boson model is studied by path measure in \cite{hhl12} and 
we can apply it in this paper 
since the Rabi model can be regarded as  the single mode photon version 
of the spin-boson model.

\subsection{Two conjectures}

Let us here consider the Rabi Hamilonian $H_{\mathrm R}$ 
before the renormalization:
$$
H_{\mathrm R}= \PF+\frac{\omega}{2}
= \s_z\Delta+\omega \left(\add a 
+\frac{1}{2}\right)+ g\s_x\left(a+\add\right).
$$ 
In this letter we follow the clasification 
proposed in \cite{Solano10}, 
and define the ultra-strong coupling regime by 
the region in which the dimensionless 
coupling strength $g/\omega>0.1$. 
 
\begin{figure}[htbp]
   \begin{center}
      {\includegraphics{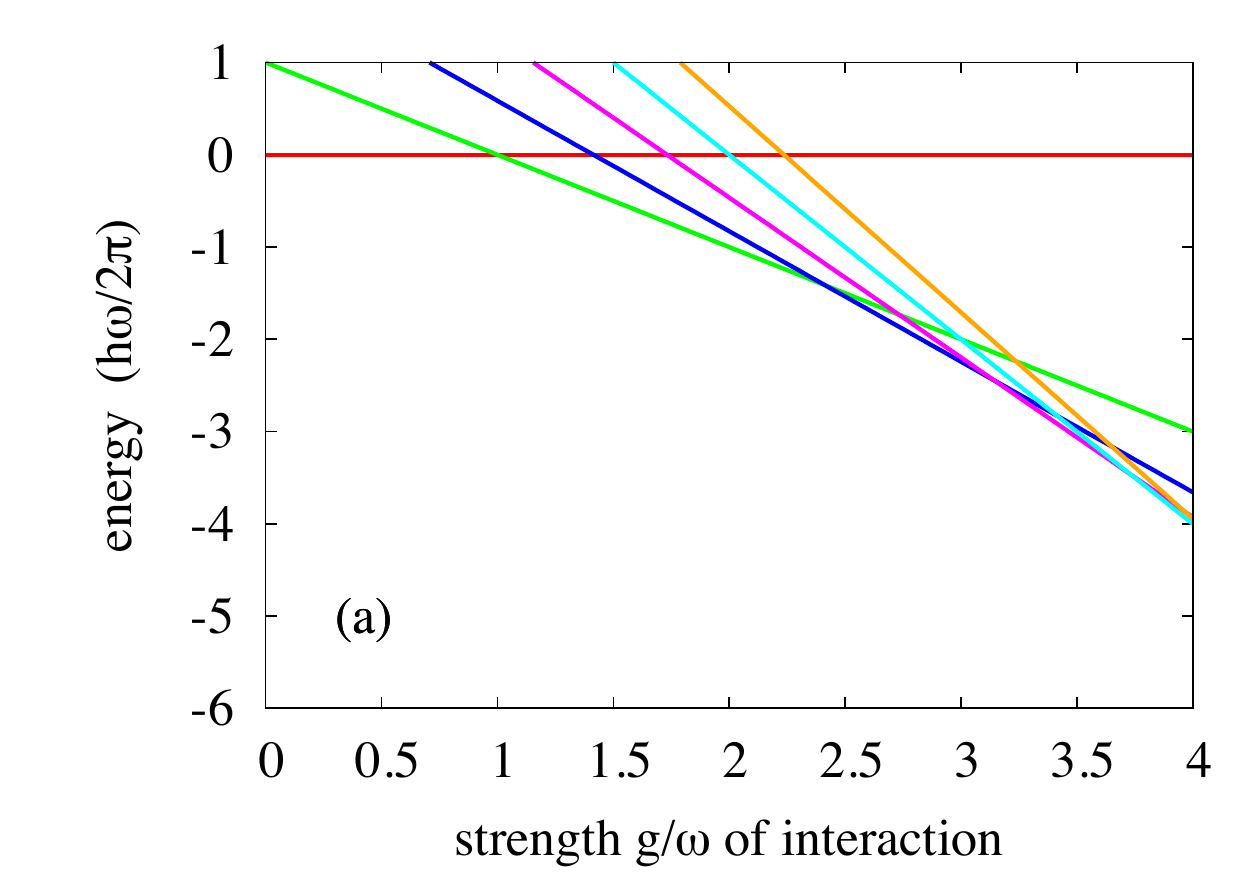}}
 \qquad 
       {\includegraphics{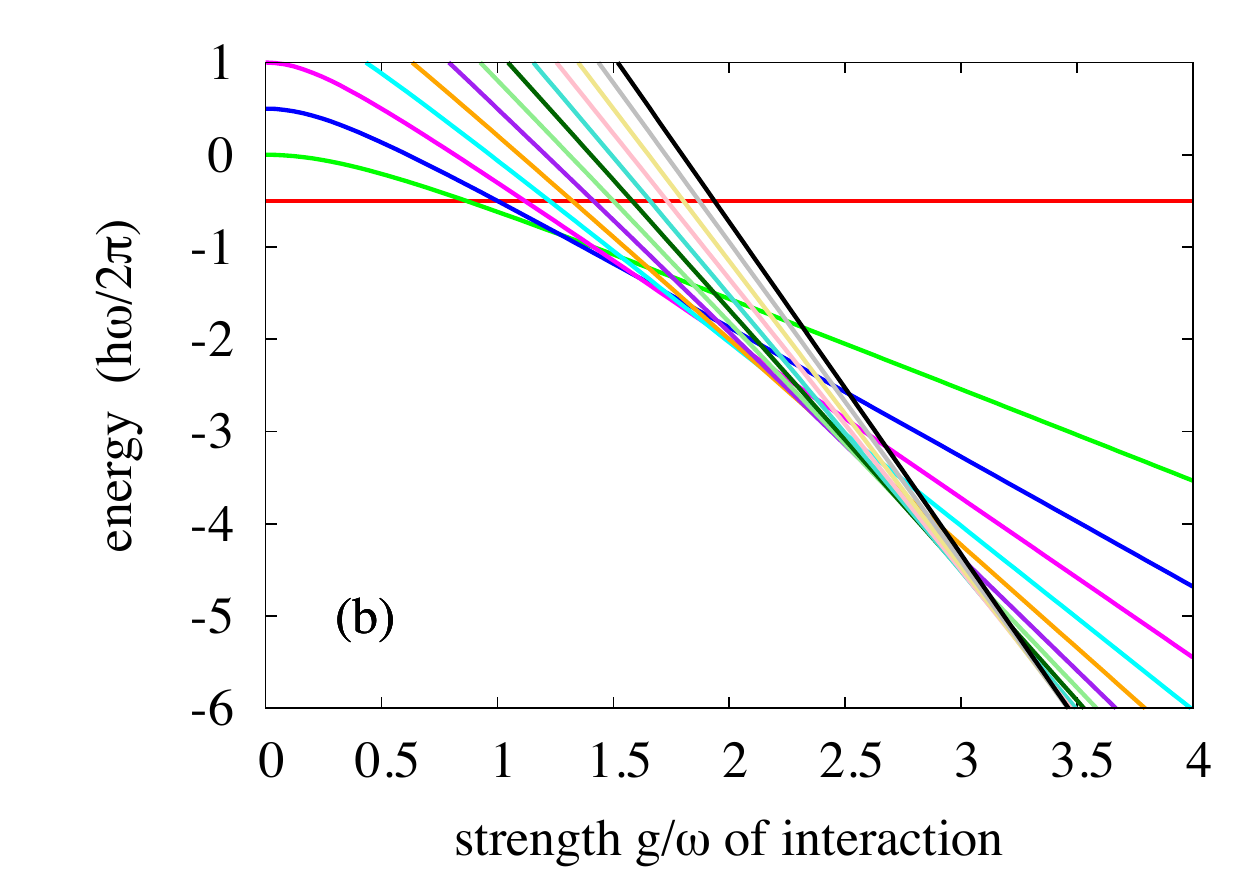}} 
   \end{center}
  \caption{\scriptsize 
Energy level crossings among $E_{\nu}^{\mathrm{JC}}(g)$, 
$\nu=0, -1, -2, \cdots$, of the JC Hamiltonian. 
Each color indicates individual index $\nu$ of the energy 
$E_{\nu}^{\mathrm{JC}}(g)$. 
(a) $2\Delta=\omega$; 
(b) $2\Delta=3\omega$.}
  \label{fig:JC} 
\end{figure}

Applying the RWA to $H_{\mathrm R}$, we have 
the JC Hasmiltonian: 
$$
H_{\mathrm{JC}}=
\s_z\Delta+\omega \left(\add a 
+\frac{1}{2}\right)+
g\left(\sigma_{-}a^{\dagger}+\sigma_{+}a\right), 
$$
where 
$\sigma_{\pm}=(\sigma_{x}\pm i\sigma_{y})/2$. 
We denote by $E_{\mathrm {JC}}$ 
the ground state energy of $H_{\mathrm{JC}}$. 
The JC model is a completely solvable model, 
and the eigenstate $\varphi_{\nu}^{\mathrm{JC}}(g)$ 
of $H_{\mathrm {JC}}$ and its corresponding eigenenergy 
$E_{\nu}^{\mathrm{JC}}(g)$ are given 
for each $\nu\in\mathbb{Z}$ in the following procedure: 
Let $\gr(x)=(\omega/\pi)^{1/4}e^{\omega x^2/2}$  be the normalized eigenvector 
associated with the lowest eigenvalue, $0$,  of 
the harmonic oscillator $\omega\add a=
\half(-\frac{d^2}{dx^2}+\omega^2 x^2-{\omega})$.
Then Fock states are defined by 
 $|n\rangle=\frac{1}{\sqrt{n!}} (\prod^n \add) \gr$ for the single mode photon, 
$n=0, 1, 2, \cdots$, with $|0\rangle=\gr$.  
For the spin ground state $|-\rangle=\vvv{0\\1}$ and 
the spin excited state $|+\rangle=\vvv{1\\0}$ of $\Delta\sigma_{z}$, 
we define states $|-,n\rangle$ and 
$|+,n\rangle$ by 
$|-,n\rangle=|-\rangle\otimes |n\rangle$ 
and  
$|+,n\rangle=|+\rangle\otimes |n\rangle$, 
respectively. 
Then, 
$$
\begin{cases}
\varphi_{0}^{\mathrm{JC}}(g)
= 
|-,0\rangle, \\  
\varphi_{+|\nu|}^{\mathrm{JC}}(g) 
=
\cos(
\theta_{|\nu|}(g))
|+,|\nu|-1\rangle 
+ \sin(
\theta_{|\nu|}(g))
|-,|\nu|\rangle,\quad \nu \ne 0, \\ 
\varphi_{-|\nu|}^{\mathrm{JC}}(g) 
= 
-\, \sin(
\theta_{|\nu|}(g))
|+,|\nu|-1\rangle 
+ \cos(
\theta_{|\nu|}(g))
|-,|\nu|\rangle, \quad \nu \ne 0, 
\end{cases}
$$
where $\theta_{|\nu|}(g)
=
\frac{1}{2}\tan^{-1}
\left(\frac{2g\sqrt{|\nu|}}{2\Delta-\omega}\right)$ 
if $2\Delta\ne\omega$; 
$\theta_{|\nu|}(g)=\pi/4$ 
if $2\Delta=\omega$, 
and 
$$
\begin{cases}
{\displaystyle 
E_{0}^{\mathrm{JC}}(g)=
\, -\, \frac{(2\Delta-\omega)}{2}}, &\quad \\ 
{\displaystyle 
E_{\pm|\nu|}^{\mathrm{JC}}(g)= 
\omega|\nu| 
\pm 
\sqrt{\frac{(2\Delta-\omega)^{2}}{4}+\mathrm{g}^{2}|\nu|\,}}, \quad 
& \nu\ne 0. 
\end{cases}
$$ 
According to \cite{hir09a,hir09b}, 
the remarkable finding for $E_{\mathrm{JC}}$ 
is the energy level crossings in the ultra-strong coupling regime: 
For each $n=0, 1, 2, \cdots$, 
there exists $g_{n+1}>0$ such that 
$E_{-n}^{\mathrm{JC}}(g)$ and $E_{-(n+1)}^{\mathrm{JC}}(g)$ 
cross each other at $g=g_{n+1}$, 
and 
$$
\begin{cases}
E_{\mathrm{JC}}=E_{-n}^{\mathrm{JC}}(g) &\text{if $g<g_{n+1}$}, \\ 
E_{\mathrm{JC}}=E_{-n}^{\mathrm{JC}}(g)=E_{-(n+1)}^{\mathrm{JC}}(g) 
& \text{if $g=g_{n+1}$}, \\ 
E_{\mathrm{JC}}=E_{-(n+1)}^{\mathrm{JC}}(g) & \text{if $g>g_{n+1}$},  
\end{cases}
$$
provided $2\Delta\ge\omega$. 
See Fig.\ref{fig:JC}. 
In other words, 
as $g$ gets large, 
there exists $\nu_{g}\in\mathbb{Z}_{-}$ 
such that $E_{\mathrm {JC}}=E_{\nu_{g}}^{\mathrm{JC}}(g)$, 
and moreover, $\nu_{g}$ is strictly decreasing 
and $\nu_{g}\to-\infty$ as $g\to\infty$. 
Namely, these energy level crossings take place and 
make the ground state energy $E_{\mathrm{JC}}$ 
as the envelop of $E_{\nu}^{\mathrm{JC}}(g)$, 
$\nu=0, -1, -2, \cdots$, in Fig.\ref{fig:JC}. 
We also note that the ground-state entanglement \cite{PZDS10} 
for the JC model. 
Namely, for instance, the ground state is a separable state 
for $g<g_{1}$, but it becomes an entangled state for $g\ge g_{1}$. 
The details on $g_{n}$ and $\nu_{g}$ are in \cite{hir09a,hir09b}. 

\begin{figure}[htbp]
  \begin{center}
     {\includegraphics{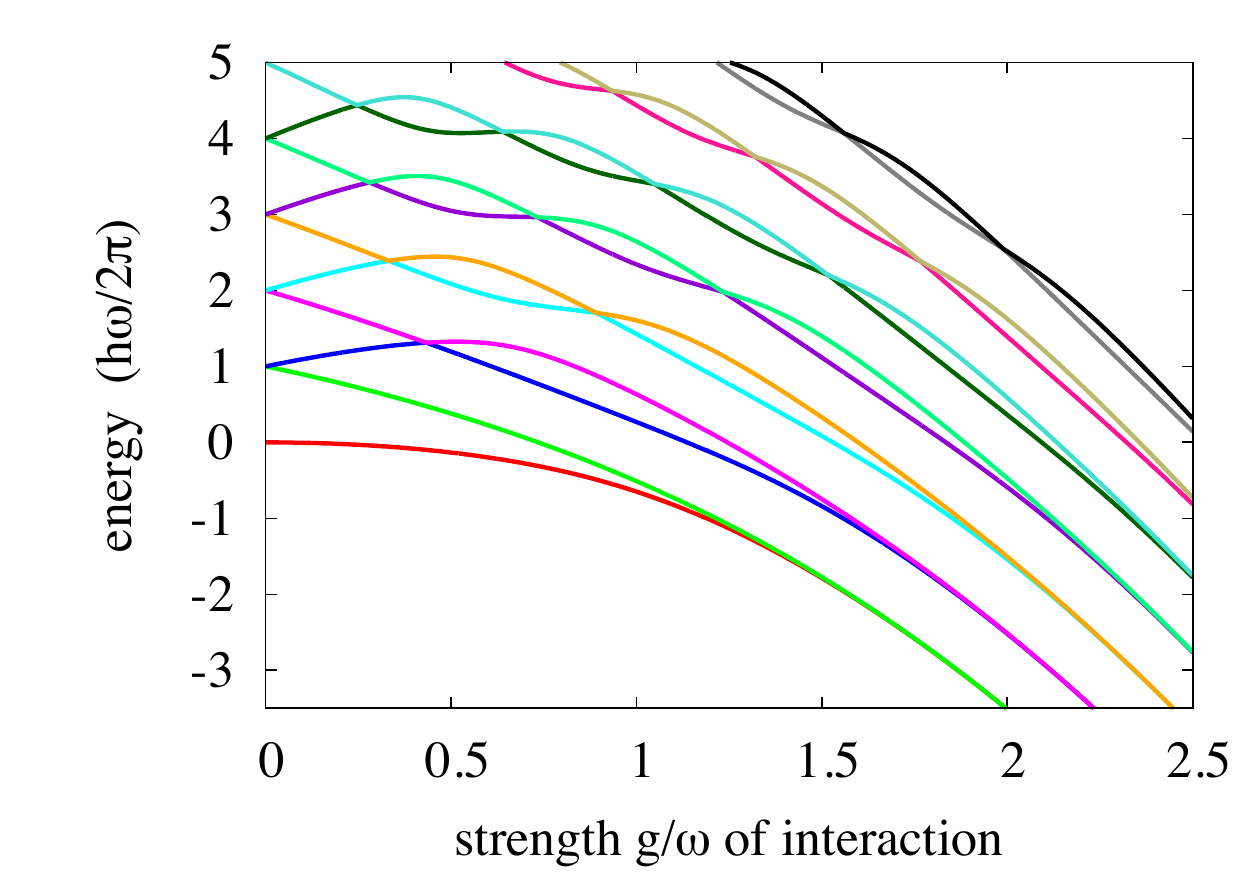}} 
  \end{center}
  \caption{\scriptsize 
Energy levels of $H_{\mathrm{R}}$ for $2\Delta=\omega$. 
Each color indicates the $n$th level of the energy 
for $n=0, 1, 2, \cdots$ from the bottom, 
where the $0$th level energy means the ground state energy.}
\label{fig:Rabi}
\end{figure}

In Fig.\ref{fig:Rabi} there is a numerical computation 
of the energy levels of $H_{\mathrm{R}}$. 
It says that 
\begin{enumerate}
\item[(C1)] there is no energy level crossing between 
the ground state energy and the $1$st excited state energy; 
\item[(C2)] we may say that there are $n$ energy level 
crossings between the $2n$-th excited state energy and 
the ($2n+1$)-th excited state energy, $n=1, 2, \cdots$. 
\end{enumerate}
In this letter, we will prove (C1).

\section{Results and proofs}
Before going to show the Feynman-Kac formula of $e^{-t\PF}$, 
we prepare a probabilistic description of $\PF$.

Let $\s=(\s_x,\s_y,\s_z)$ be elements of $SU(2)$. 
The rotation group in $\RR^3$ 
has an adjoint representation on $SU(2)$.
Let $n\in \RR^3$ be a unit vector and $\theta\in [0,2\pi)$. Thus we have 
$e^{(i/2)\theta n\cdot \s}$ satisfies that 
$e^{(i/2)\theta n\cdot \s}
\s_\mu e^{-(i/2)\theta n\cdot \s}=(R\s)_\mu$, where 
$R$ denotes $3\times 3$ matrix representing  
the rotation around $n$ with angle $\theta$.
In particular for $n=(0,1,0)$ and $\theta=\pi/2$, 
we have 
$
e^{(i/2)\theta n\cdot \s}
\s_x e^{-(i/2)\theta n\cdot \s}=\s_z
$ and 
$
e^{(i/2)\theta n\cdot \s}
\s_z e^{-(i/2)\theta n\cdot \s}=-\s_x$.
Set $U=e^{(i\pi/4)\s_y}
$. Then 
\eq{rabi2}
U \PF U^{-1}= 
 \omega \add a +g\s_z(a+\add)- \s_x\Delta.
\en
Since $\gr$  is strictly positive, we can  define the unitary operator 
$U_g:\LR\to L^2(\RR, \gr^2 dx)$ by 
$U_g f=\gr^{-1} f$. 
We set the  probability measure $\gr^2 dx$ on $\RR$ by $d\mu$. Thus $U \PF U^{-1}$ is transformed to 
the operator: 
\eq{rabi3}
U_g U \PF U^{-1}U_g^{-1}=
\half \lk -\frac{d^2}{dx^2}+\omega x\frac{d}{dx}\rk   +g\s_z\sqrt{2\omega} x - \s_x\Delta
\en
 in $\CC^2\otimes L^2(\RR,d\mu)$. 
 Let us introduce 
 $\zz=\{-1,+1\}$ to redefine the Hamiltonian \kak{rabi3} on  a set of {\it scalar} functions.  
 We identify $\CC^2\otimes \LRM$ with 
\eq{hilbertspance}
{\ms H}={L^2(\RR\times\zz,d\mu)}=
\lkk 
f=f(x,\s)\left|
\sum_{\s\in\zz} \int\!\!  |f(x,\s)|^2d\mu(x)<\infty
\right.\rkk
\en 
by 
$\CC^2\otimes \LRM\ni 
\vvv{f_+(x)\\ f_-(x)}\mapsto f(x,\s)\in {\ms H}$.
Thus 
under this identification 
\kak{rabi3}
is transformed to 
the operator $H$: 
\eq{hamiltonian}
Hf(x,\s)=
\lkk
\half\left(
-\frac{d^2}{dx^2}+\omega x\frac{d}{dx}
\right)  
+g \sqrt{2\omega} \s x\rkk f(x, \s) - \Delta f(x,-\s),\quad \s\in\zz
\en
in $\ms H$. 
Thus we have the lemma below:
\bl{unitary}
The operator 
$\PF$ in $\CC^2\otimes\LR$ is unitarily equivalent to 
$H$ in $\ms H$.
\el
In what follows we deal with $H$ instead of $\PF$. 
Let 
$$h=\half\lk  -\frac{d^2}{dx^2}+\omega x\frac{d}{dx}\rk   
$$ and $\pro X$ be the Ornstein-Uhrenbeck process on some probability space 
$(C, \Sigma, P^x)$.
We have ${P^x}(X_0=x)=1$
\begin{align*}
&\int\!\!  d\mu(x)  \Ebb_{P^x} \lkkk X_t \rkkk =0,\\
&
\int\!\!  d\mu(x)  \Ebb_{P^x} \lkkk X_t X_s\rkkk =\frac{e^{-|t-s|\omega}}{2\omega}.
\end{align*}
Here $\Ebb_Q\lkkk \cdots\rkkk $ denotes the expectation with respect to a probability measure $Q$. 
The generator of $X_t$ is given by $-h$ and 
$$(f, e^{-th}g)_{\ms H}=\int\!\!  d\mu(x)  \Ebb_{P^x}\lkkk \ov{f(X_0)} g(X_t)\rkkk .$$
The distribution $\rho_t(x,y)$ of $X_t$ under ${P^x}$ is given by 
\eq{distribution}
\rho_t(x,y)=\gr(x)^{-1} K_t(x,y) \gr(y),
\en
where $K_t(x,y)$ denotes the Mehler kernel given by 
$$K_t(x,y)=\frac{1}{\sqrt{\pi(1-e^{-2t})}}\exp\lk
\frac{4xy e^{-t}-(x^2+y^2)(1+e^{-2t})}{2(1-e^{-2t})}\rk.
$$
See e.g., \cite[3.10.4]{lhb11} for the detail  of Ornstein-Uhrenbeck processes and harmonic oscillators.  
In order to show the spin part by a path measure we introduce 
a Poisson process. 
Let  $\pro N$ be a Poisson process on some probability space $(C',\Sigma',\nu)$ with unit intensity, i.e., 
$$\Ebb_\nu\lkkk \one_{N_t=n}\rkkk =\frac{t^n}{n!} e^{-t},\quad n\geq 0.$$
We define 
$\s_t=(-1)^{N_t}\s $, $\s\in\zz$,  for $t\geq 0$.
Let 
\eq{sum}
\sum_{\s\in\zz}\int\!\!  d\mu(x)\Ebb_{P^x}\Ebb_\nu\lkkk \cdots\rkkk =\Ebb\lkkk \cdots\rkkk .
\en
\bt{FKF}
It follows that 
\begin{align}
\label{fkf}
(\Delta>0)\ \ \ &
(f, e^{-tH}g)_{\ms H}=
e^t
\Ebb
\lkkk 
\ov{f(X_0,\s_0)}g(X_t,\s_t) e^{-g\sqrt{2\omega}\int _0^t \!\! \s_sX_sds}\Delta^{N_t}
\rkkk,\\
\label{fkf2}
(\Delta=0)\ \ \ &
(f, e^{-tH}g)_{\ms H}=
e^t
\Ebb
\lkkk \one_{N_t=0}\ov{f(X_0,\s)}g(X_t,\s) e^{-g\s \sqrt{2\omega}\int _0^t \!\! X_sds}\rkkk 
\end{align}
\et
\proof 
Let $\Delta>0$. 
By a minor modification of \cite[Theorem 5.10]{hil12} we can see that 
\eq{fkf1}
(f, e^{-tH}g)_{\ms H}
=e^t
\Ebb\lkkk 
\ov{f(X_0,\s_0)}g(X_t,\s_t) e^{-g\sqrt{2\omega}\int _0^t \!\! \s_sX_sds}
e^{\int _0^t \!\! \log \Delta dN_s}
\rkkk .\en
Here 
$\d \int  _0^t \!\! f(N_s) dN_t=\sum_{r, N_{r+}\not= N_{r-}}f(N_r)$.
Since 
$e^{\int _0^t \!\! \log \Delta dN_s}=e^{\log \Delta^{N_t}}=\Delta^{N_t}$, 
\kak{fkf} follows.
In the case of $\Delta=0$ 
only the set $N_t=0$ contributes to the path integral. 
Then
\eq{fkf11}
(f, e^{-tH}g)_{\ms H}
=e^t
\Ebb
\lkkk \ov{f(X_0,\s_0)}g(X_t,\s_t) e^{-g\sqrt{2\omega}\int _0^t \!\!\s_sX_sds}
\one_{N_t=0}
\rkkk .\en
Then \kak{fkf2} follows. 
\qed
Let $E=\is(H)$.
\bc{uniqueness}
It follows that ${\rm dim}{\rm ker}(H-E)=1$, i.e., the ground state of $\PF$ is unique. 
\ec
\proof 
Let $f,g\geq0$ but not identically zero.
Then for sufficiently small $\epsilon>0$, 
we see that both 
$\Omega_f=\{(x,\s)\in\RR\times\zz|f(x,\s)>\epsilon\}$
and 
$\Omega_g=\{(x,\s)\in\RR\times\zz|g(x,\s)>\epsilon\}$
have positive measures.
We  have 
 by \kak{fkf},  
$$(f, e^{-tH}g)\geq\epsilon e^t 
\Ebb\lkkk \one_{\Omega_f}(X_0,\s_0)\one_{\Omega_g}(X_t,\s_t) e^{-g\sqrt{2\omega}\int _0^t \!\!\s_sX_sds}\Delta^{N_t}\rkkk .$$
Since $\Omega_f$ is a subset of $\RR\times \zz$, 
we have $\Omega_f=\bigcup_{\s\in\zz}(\Omega_f^\s,\s)$. 
Thus either 
$\Omega_f^+$ or $\Omega_f^-$ ($\subset \RR$) have at least 
a positive measure. Suppose that $\Omega_f^+$ has a positive measure. 
Similarly we see that $\Omega_g=\bigcup_{\s\in\zz}(\Omega_g^\s,\s)$ and suppose that $\Omega_g^+$ is a positive measure. 
Let $\Omega$ be the set of paths starting from  
the inside of $(\Omega_f^+,+)$ 
and arriving at the inside of $(\Omega_g^+,+)$. 
We see that 
\begin{align*}
\Ebb
\lkkk \one_\Omega\rkkk &= 
\Ebb
\lkkk \one_{\Omega_f^+}(X_0)\one_{\Omega_g^+}(X_t)\one_{N_t={\rm even}}\rkkk .
\end{align*}
By using the distribution $\rho_t$ of $X_t$ we have 
\begin{align*}
\Ebb\lkkk \one_\Omega\rkkk 
= \sum_{n=0}^\infty \frac{t^{2n}}{(2n)!} e^{-t}\int _{\Omega_f^+} \!\!dx
\int _{\Omega_g^+}\!\!dy \gr(x) K_t(x,y) \gr(y)
>0.
\end{align*}
Hence  we conclude that 
$\Omega$ has a positive measure and 
$$(f, e^{-tH}g)\geq\epsilon e^t 
\Ebb
 \lkkk \one_{\Omega} e^{-g\sqrt{2\omega}\int _0^t \!\!\s_sX_sds}\Delta^{N_t}\rkkk >0.$$
Thus $e^{-tH}$ is a positivity improving operator. 
Thus ${\rm dim}{\rm ker}(H-E)=1$ follows from the Perron-Frobenius theorem. 
\qed
\bc{crossing}
The ground state energy  of $\PF$ has no crossing for all the values of $g$ and $\Delta$. 
\ec

\section{Conclusion} 

In this letter we have proved the first conjecture (C1) 
that the numerical computation predicts, 
while the JC model has many energy level crossings 
for the ground state energy 
in the ultra-strong coupling regime of circuit QED 
though it has no energy level crossing 
in the weak and strong coupling regimes.  
It shows that it is in the ultra-strong coupling regime 
that there is a big qualitative difference 
as well as quantitative one between the JC model 
and the Rabi model.

\qquad

\noindent
{\bf Acknowledgments:} This work is financially supported by 
Grant-in-Aid for Science Research (B) 20340032
from JSPS. 
One of authors (M.H.) acknowledges the support from JSPS, 
Grant-in-Aid for Scientific Research (C) 23540204, 
and he also expresses special thanks to 
Pierre-Marie Billangeon and Yasunobu Nakamura 
for useful discussions, which aroused his interest 
in circuit QED.

\end{document}